\documentclass[twocolumn,nofootinbib,amsmath,amssymb]{revtex4}

\usepackage{graphicx}
\usepackage{dcolumn}
\usepackage{bm}
\usepackage{afterpage}

\newcommand{\etal}{{\it et al.}}
\begin{document}

\title{Exclusive Semileptonic $b\to u\ell\nu$ Decays at CLEO}

\author{S. Stone}
 \email{stone@physics.syr.edu}
\affiliation{Physics Department of Syracuse University\\
Syracuse NY, 13104, USA\\
}%

\begin{abstract}
Updated CLEO results are presented for branching ratios and
four-momentum transfer, $q^2$, dependence of exclusive charmless
semileptonic $B$ decays, where the final state hadron is either a
$\pi$, $\rho$, $\omega$, $\eta$ or $\eta'$. These results have
comparable accuracies with those of other experiments. We address
the issue of flavor singlet couplings by limiting
$\Gamma(B^+\to\eta'\ell^+\nu)/\Gamma(B^+\to\eta\ell^+\nu)>
2.5~{\rm at~90\%~CL}$. We also extract a value of
$|V_{ub}|=\left(4.3\pm 0.4 \pm 0.2^{+0.6}_{-0.4}\right)\times
10^{-3}$ in one particular unquenched lattice QCD model using
$\pi\ell^+\nu$ data above $q^2$ of 16 GeV$^2$.
\end{abstract}

\maketitle

\section{Introduction}
In this paper I present improved measurements of $B^0\to
\pi^-\ell^+\nu$ and $\rho^-\ell^+\nu$ branching ratios and
four-momentum transfer, $q^2$, dependencies. Also shown is evidence
for  ${\cal{B}}(B^+\to \eta' \ell^+\nu$) and an upper limit for
$B^+\to \eta\ell^+\nu$. These result use a ``neutrino
reconstruction" technique that is an extension of a previous CLEO
analyses \cite{athar}, based on CLEO II and II.V data. Here we
include CLEO III data, representing an increase of 60\% in the
sample, for a total of 15.4$\times 10^6$ $B\overline{B}$ events.

Semileptonic $B$ decays proceed when the $b$ quark transforms to a
$c$ or $u$ quark emitting a virtual $W$ that manifests itself as a
lepton-neutrino pair. For a $B$ meson decaying into a single hadron
($h$), the decay rate can be written exactly in terms of the
four-momentum transfer defined as:
\begin{equation}
q^2=\left(p^{\mu}_B-p^{\mu}_h\right)^2=m_B^2+m_h^2-2E_hm_B~.
\end{equation}
For decays to light pseudoscalar hadrons, via the $b\to u$
transition, and ``virtually massless" leptons, the decay width is
given by:
\begin{equation}
{{d\Gamma(B\to P \ell^+\nu)}\over{dq^2}}
={{|V_{ub}|^2G_F^2p_P^3}\over{24\pi^3}}\left|f_+(q^2)\right|~,
\end{equation}
where $p_P$ is the three-momentum of $P$ in the $D$ rest frame, and
$f_+(q^2)$ is a ``form-factor," whose normalization must be
calculated theoretically, although its shape can be measured, in
principle.

For $B$ decays into a vector meson final state we measure both $q^2$
and $\cos \theta_{WL}$, where $\theta_{WL}$ is the angle between the
$\ell^+$ direction in the $W$ rest frame and the $W$ direction in
the $B$ rest frame. The double differential decay rate then is
related to the helicity amplitudes, $H_i$ as
\begin{eqnarray}
&&\frac{d\Gamma\left(B\to V\ell^+\nu\right)}{dq^2d\cos
\theta_{WL}}=\\\nonumber &
&~~~~~~~~~~\left|V_{ub}\right|^2\frac{G_F^2M_B^2P_Vq^2}{128\pi^3}\left[(1-\cos\theta_{WL})^2\frac{|H_+|^2}{2}\right.\\\nonumber
& &~~~~~~~~~~\left.+(1+\cos\theta_{WL})^2\frac{|H_-|^2}{2}+\sin^2
\theta_{WL}|H_0|^2\right].
\end{eqnarray}

\section{Experimental Methods}
We first select events with either an $e^{\pm}$ or $\mu^{\pm}$ with
momenta greater than 1 GeV/c. Events with more than one such lepton
are rejected, since multiple leptons are indicative of more than one
semileptonic decay. We also require the events to be of spherical
shape in momentum space, and to have a net observed charge of zero
except in modes with a single pseudoscalar hadron where the
requirement is loosened to $\pm$1.

The technique of neutrino reconstruction makes use of the energy and
momentum deposited by all found (i. e. visible) charged tracks and
photons in the event. Then the neutrino four-vector is formed from
the missing energy and momentum as:
\begin{eqnarray}
\overrightarrow{p}_{\rm miss}&=&\overrightarrow{p}_{\!\rm
CM}-\overrightarrow{p}_{\!\rm visible}\\\nonumber E_{\rm
miss}&=&E_{\rm CM}-E_{\rm visible}.
\end{eqnarray}
The resolution in $\overrightarrow{p}_{\!\rm miss}$ is $\sim$0.1
GeV/c, r.m.s. Since the missing-mass-squared
\begin{equation}
{\rm MM}^2=E^2_{\rm miss}-\overrightarrow{p}_{\!\rm miss}^2,
\end{equation}
should peak at zero, we require MM$^2/2E_{\rm miss}<0.5$ GeV. This
requirement takes into account the scaling of the MM$^2$ resolution
proportional to $E_{\rm miss}$.

Then we look at signals in the $\Delta E$-invariant mass plane
($M_{\rm h\ell\nu}$), where
\begin{eqnarray}
\Delta E &=&E_{\rm h}+E_{\ell}+E_{\rm miss}-E_{\rm beam}\\\nonumber
M_{\rm h\ell\nu}^2&=&E_{\rm beam}^2-\left(\overrightarrow{p}_{\rm
h}^2+\overrightarrow{p}_{\!\ell}^2+\alpha\overrightarrow{p}_{\!\rm
miss}^2\right);
\end{eqnarray}
here $E_{\rm miss}$ is replaced with $|p_{\rm miss}|$, because the
latter has better resolution, and the parameter $\alpha$ is adjusted
for each hypothesis in order ensure that $\Delta E$ is zero.

\begin{figure}[htb]
\centerline{
\includegraphics[width=7.cm]{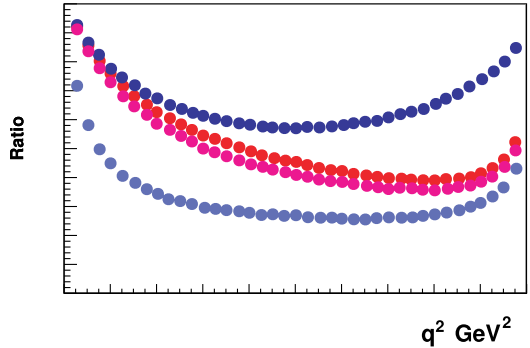}
} \vspace{-.3mm}\caption{Predictions of different models for the
ratio of form-factors for $\cos\theta_{WL}<0/\cos\theta_{WL}>0$, as
a function of $q^2$. From top to bottom \cite{Mel, Ball, ukqcd,
ISGW2}. } \label{ffwl}
\end{figure}

The $\pi^-\ell^+\nu$ data are analyzed in four separate bins of
$q^2$. For $\rho\ell^+\nu$, the models differ greatly on their
$\cos\theta_{WL}$ dependencies. Fig.~\ref{ffwl} shows the ratio of
the predicted form-factors for
$\cos\theta_{WL}<0/\cos\theta_{WL}>0$, as a function of $q^2$. These
differences can lead to substantial systematic errors. We reduce
these by analyzing the data in two separate $\cos\theta_{WL}$ bins,
one above and one below zero and four $q^2$ bins for a total of 8
intervals.

The $\eta'\ell^+\nu$ data are split into two $q^2$ bins at 10
GeV$^2$, while the $\eta\ell^+\nu$ sample is integrated over all
$q^2$.

\begin{figure}[htb]
\centerline{
\includegraphics[width=4.2cm]{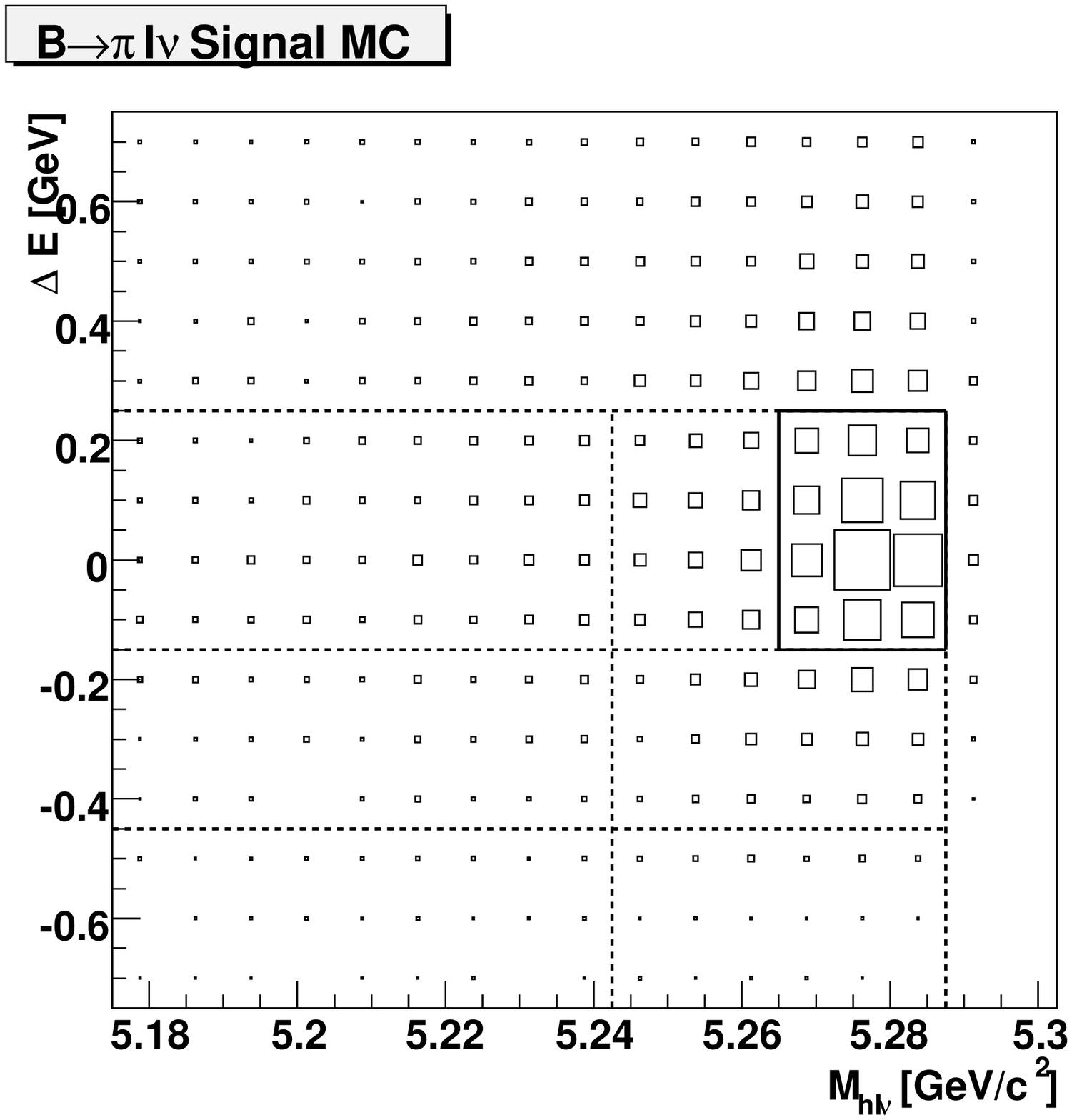}
\includegraphics[width=4.2cm]{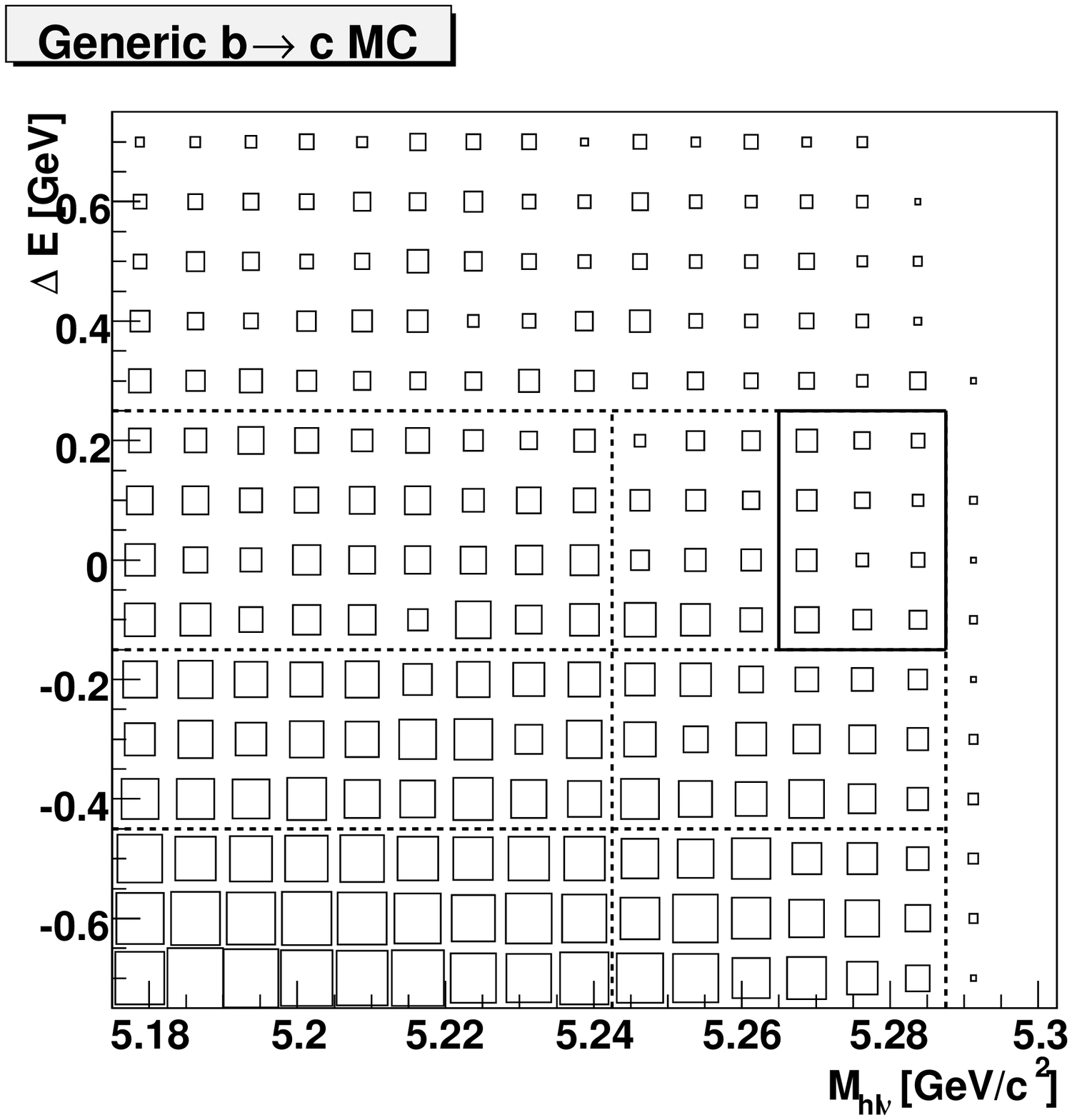}
} \vspace{-.2mm}\caption{Distribution of $B^0\to\pi^-\ell^+\nu$
(left) and generic $b\to c$ events (right) in variables $M_{\rm
h\ell\nu}$ and $\Delta E$ from Monte Carlo simulation. The signal
region is shown by the rectangular box.} \label{sigback}
\end{figure}

Monte Carlo simulation of signal semileptonic events and generic
$b\to c$ backgrounds in terms of the kinematic variables $M_{\rm
h\ell\nu}$ and $\Delta E$ is shown in Fig.~\ref{sigback}. The
backgrounds peak at low values of $M_{\rm h\ell\nu}$ and $\Delta E$,
and are well separated from the signal which peaks at the $B$ mass
and $\Delta E$ of zero.

\begin{figure}[htbp]
\centerline{
\includegraphics[width=8.5cm]{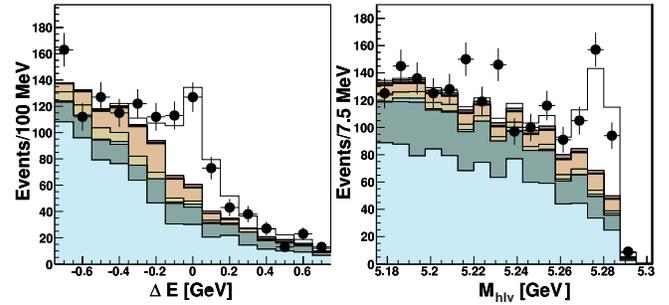}
} \vspace{-.2mm}\caption{Projections of the data fit for $(\pi^- +
\pi^0)\ell^+\nu$, shown as points with error bars. The background
components are (listed from bottom to top in different shades):
$B\to X_c \ell^+\nu$, continuum, other $B\to X_u \ell^+\nu$
channels, specific $B\to (\rho+\omega) \ell^+\nu$ cross-feed,
$\pi^+$ and $\pi^0$ cross-feed. The solid line shows the sum. }
\label{pi-ellnu}
\end{figure}

\begin{figure}[htb]
\centerline{
\includegraphics[width=8.5cm]{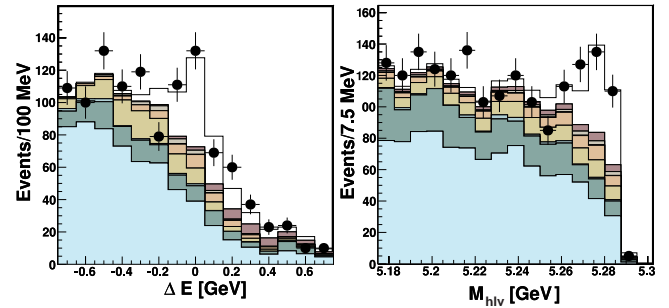}
} \vspace{-.2mm}\caption{Projections of the data fit for $(\rho^+
+\rho^0 + \omega)\ell^+\nu$, shown as points with error bars. The
background components are the same as in Fig.~\ref{pi-ellnu}, except
that the $\rho\ell^+\nu$ background is $\pi\ell^+\nu$. The solid
line shows the sum. } \label{rho-ellnu}
\end{figure}

\begin{figure}[htb]
\centerline{
\includegraphics[width=8.5cm]{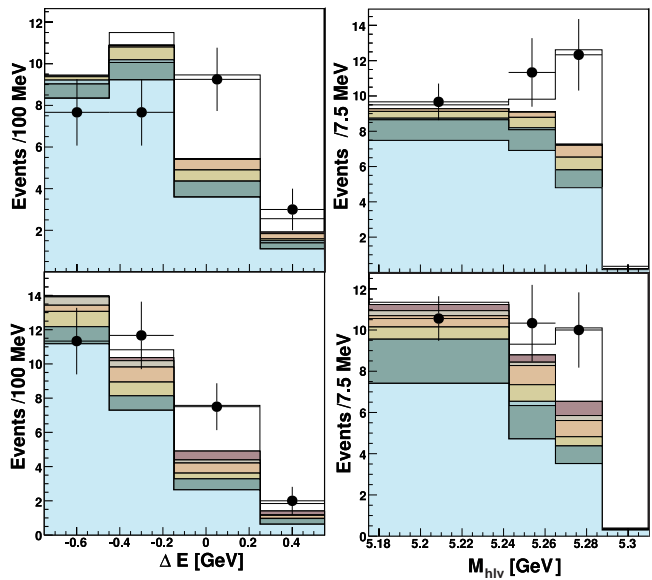}}
 \vspace{-.2mm}\caption{The projections of the data fit
for $\eta'\ell^+\nu$ (top) and $\eta\ell^+\nu$ (bottom), shown as
points with error bars. The background components are the same as in
Fig.~\ref{pi-ellnu}. The solid line shows the sum. }
\label{etap-ellnu}
\end{figure}

The data are fit simultaneous for all modes in the separate bins of
$q^2$ and $\cos\theta_{WL}$ discussed above. The projections of the
fits for $\pi\ell^+\nu$, $(\rho+\omega)\ell^+\nu$ and
$\eta^{(\prime)}\ell^+\nu$ are shown in Figs.~\ref{pi-ellnu},
\ref{rho-ellnu} and \ref{etap-ellnu}, respectively. For the $\eta'$
final state only data for $q^2<10$ GeV$^2$ is used, while the other
modes are shown summed over all $q^2$.

 The branching fractions, integrated over
$q^2$ are given in Table~\ref{tab:bfs}. The $\eta'\ell^+\nu$ is
observed at the 3$\sigma$ level, while for $\eta\ell^+\nu$ we have a
substantially smaller upper limit. Thus we can quote the ratio
\begin{equation}
R'\equiv\frac{\Gamma(B^+\to\eta'\ell^+\nu)}{\Gamma(B^+\to\eta\ell^+\nu)}>
2.5~{\rm at~90\%~CL}.
\end{equation}
Anomalously large inclusive production of $\eta'$ at high momentum
has been observed in $B$ decays \cite{CLEOetap}. Several models have
attempted to explain this phenomena by an enhanced gluonic
form-factor \cite{Ali}, or by flavor singlet coupling
\cite{singlet}. Measurement of inclusive $\eta'$ production from the
$\Upsilon(1S)$ has ruled out the form-factor explanation
\cite{CLEO1S}. Our limit on $R'$ supports an enhanced flavor singlet
coupling.

The systematic errors in these measurements are dominated by the
uncertainty on neutrino reconstruction, which is particular large in
the $\eta$ and $\eta'$ modes due to shower resolution.

\begin{table}
\caption{Measured branching fractions. The modes with $\pi^0$ or
$\rho^0$ and $\omega^0$ are averaged in assuming isospin symmetry.
Errors are (in order): statistical, systematic and model.
\label{tab:bfs}}
\begin{ruledtabular}
\begin{tabular}{lc}
    Final State  &  ${\cal{B}}\times 10^{-4}$       \\
$\pi^-\ell^+\nu$ & $1.37\pm 0.15\pm0.12\pm0.01$\\
$\rho^-\ell^+\nu$ & $2.93\pm 0.37\pm0.39\pm0.04$\\
$\eta\ell^+\nu$ & $<1.01$ at 90\% CL\\
$\eta'\ell^+\nu$ &$2.66\pm 0.80\pm0.57\pm0.04$\\
\end{tabular}
\end{ruledtabular}
\end{table}

These branching fractions are of comparable precision to other
results, some of which used much larger data samples. They have been
tabulated by the Heavy Flavor Averaging Group \cite{HFAG}.

\begin{figure}[tb]
\vskip -2mm \centerline{
\includegraphics[width=9cm]{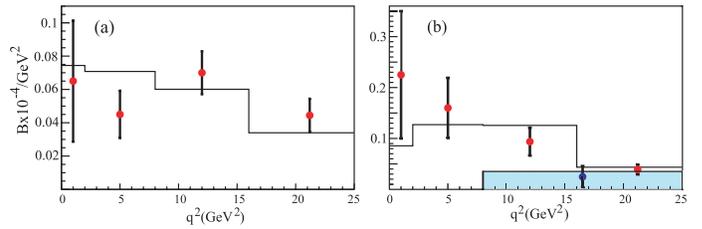}
} \vspace{-1mm}\caption{Branching fractions as a function of $q^2$
for (a) $\pi\ell^+\nu$ and (b) $\rho\ell^+\nu$ shown as points with
error bars. In (a) the solid lines indicate the HPQCD prediction
\cite{HPQCD} normalized to the data, and in (b) the model of Ball
and Zwicky \cite{Ball}. In (b) the shaded point is for
$\cos\theta_{WL} <0$, while the other points are for
$\cos\theta_{WL} >0$. } \label{pi-rho_br_results}
\end{figure}

 The results for the $q^2$ dependence of the branching ratios are
shown in Fig.~\ref{pi-rho_br_results} and compared with two
theoretical predictions.

We can translate these measurements into values of $|V_{ub}|$ using
theoretical models. For example, using our $B\to \pi\ell^+\nu$  rate
in the $q^2 > 16$ GeV$^2$ range, and the HPQCD unquenched lattice
QCD calculation \cite{HPQCD}, results in
\begin{equation}
|V_{ub}|=\left(4.3\pm 0.4 \pm 0.2^{+0.6}_{-0.4}\right)\times
10^{-3},
\end{equation}
where the last error is due to the model. This results are
approximately a factor of two less precise statistically than the
current $b$-factory results \cite{PDG}.

\section{Conclusions}
We find ${\cal{B}}(B^0\to\pi^-\ell^+\nu)=(1.37\pm
0.15\pm0.12\pm0.01)\times 10^{-4}$ and
${\cal{B}}(B^0\to\rho^-\ell^+\nu)= (2.93\pm
0.37\pm0.39\pm0.04)\times 10^{-4}$. The model dependent systematic
errors have been greatly reduced by determining the partial
branching ratios in bins of $q^2$ and $\cos\theta_{WL}$.

 We
extract a value of  $|V_{ub}|=\left(4.3\pm 0.4 \pm
0.2^{+0.6}_{-0.4}\right)\times 10^{-3}$ using the unquenched lattice
QCD model of HPQCD with our $\pi\ell^+\nu$ data above $q^2$ of 16
GeV$^2$.

We also show that $\eta'\ell^+\nu$ is more than 2.5 times larger
than $\eta\ell^+\nu$ leading credence to an enhanced flavor singlet
coupling.

\begin{acknowledgments}
I thank the U. S. National Science Foundation for support. I had
useful conversations concerning this work with M. Artuso, and R.
Gray.
\end{acknowledgments}

\end{document}